\documentclass[preprint]{aastex}
\usepackage{graphicx}
\usepackage{epsfig}
\usepackage{enumerate}
\def\beq{\begin{equation}}
\def\eeqno#1{\label{#1}\end{equation}}

\def\rarrow{\rightarrow }

\def\dleft{\rlap{{\it D}}\raise 8pt
\hbox{$\scriptscriptstyle\Leftarrow$}}
\def\dright{\rlap{{\it
D}}\raise 8pt\hbox{$\scriptscriptstyle\Rightarrow$}}

\def\az{a_{0}}

\def\l0{\ell_{0}}

\def\s{\sigma}

\def\l{\lambda}

\def\m{\mu}

\def\z{\zeta}

\def\xlimin{{x\rarrow\infty \atop{\raise 1pt\hbox to 30pt
{\rightarrowfill}}}}
\def\limlim#1#2{{#1\rarrow #2 \atop{\raise 1pt\hbox to 30pt
{\rightarrowfill}}}}

\def\m{\mu}

\def\Up{\Upsilon}

\def\_#1{_{\scriptscriptstyle #1}}
\def\^#1{^{\scriptscriptstyle #1}}

\begin{document}
\title{Critical take on ``Mass models of disk galaxies from the DiskMass Survey in MOND''}
\author{Mordehai Milgrom}
\affil{Department of Particle Physics and Astrophysics, Weizmann Institute}

\begin{abstract}
Angus et al. (2015) have recently faulted MOND as follows: Studying thirty disc galaxies from the DiskMass survey,
they derive the profiles of velocity dispersion perpendicular to the discs as predicted by MOND, call them $\s\_M(r)$. These are then compared with the dispersion profiles, $\s(r)$, measured as part of the DiskMass project. This is a new (theory dependent) test of MOND, different from rotation-curve analysis.

{\it A nontrivial accomplishment of} MOND -- not discussed by Angus et al. -- is that the shapes of $\s\_M$ and $\s$ agree very well, i.e $\eta(r)\equiv\s\_M(r)/\s(r)$ is well consistent with being $r$-independent (while $\s$ and $\s\_M$ are strongly $r$ dependent). The fault found with MOND was that $\eta$ is systematically above 1 (with an average of about 1.3).

I have suggested to Angus et al. that the fault may lie with the DiskMass dispersions, which may well be too low for the purpose at hand: Being based on population-integrated line profiles, they may be overweighed by younger populations, known to have much smaller dispersions, and scale heights, than the older populations, which weigh more heavily on the light distributions.

I discuss independent evidence that supports this view, and show, besides, that if the DiskMas dispersions are underestimates by only 25\%, on average, the MOND predictions are in full agreement with the data, in shape and magnitude.

Now, Aniyan et al. (2015) have questioned the DiskMass $\s$ on the same basis. They show for the solar column in the Milky Way that: `` Combining the (single) measured velocity dispersion of the total young + old disc population... with the scale height estimated for the older population would underestimate the disc surface density by a factor of $\sim 2.$'' Or, equivalently, that the population-integrated dispersion underestimates the proper $\s$ by $\sim 30\%$.
If this mismatch found for the Milky Way is typical, correcting for it would bring the measured DiskMass $\s(r)$ to a remarkable agreement with the predicted MOND $\s\_M(r)$.

\end{abstract}
\maketitle
\section{Introduction}

Angus et al. (2015) have recently considered a novel test of MOND.\footnote{For reviews of the MOND alternative to dark matter see, e.g., Famaey \& McGaugh (2012) and Milgrom (2014a).} They used thirty disc galaxies studied by the DiskMass survey (Bershady et al. 2010a, 2010b), which measured rotation curves, surface-density profiles for gas, and brightness distributions for stars, as well as profiles of velocity dispersions vertical to the galactic disc, $\s(r)$.
For these galaxies, Angus et al. performed a rotation-curve analysis based on MOND,\footnote{The DiskMass galaxies are anything but ideal for rotation-curve analysis, since to serve the purposes of the project, they were chosen to have low inclinations $i\lesssim 45^o$.} fitting for the best value of the K-band, stellar mass-to-light ratio, $\Up\_K=M/L\_K$. These $\Up\_K$ values agree with those determined from population-synthesis, as was also shown in earlier MOND rotation-curves studies (see, e.g., Figure 28 of Famaey \& McGaugh 2012).
Then, using these $\Up\_K$ values, and the DiskMass light and gas distributions, they calculated the predicted MOND profiles of the vertical velocity dispersions, call them $\s\_M(r)$.
\par
This was done numerically, using the QUMOND version of nonrelativistic MOND (Milgrom 2010). We can schematically summarize the procedure they use by a relation of the form
\beq \s^2\_M(r)=q(r)h_z\Sigma_b(r), \eeqno{i}
which captures the correct interrelations between quantities.
Here, $\Sigma_b(r)$ is the baryonic disc surface density, gotten from the K-band light surface density combined with the above MOND, best-fit $\Up\_K$ (plus the gas contribution), $h_z$ is ``the'' scale height, assumed $r$-independent, and not directly measured, but determined from $h_z-h\_R$ correlations deduced from the photometry of (other) edge-on galaxies ($h\_R$ is the measured scale length). The $r$-dependent factor $q(r)=f\z\_M(r)$, where $f$ is the constant that would appear in Newtonian dynamics ($f\sim 2\pi G$ depends on the velocity distribution), and, $\z\_M(r)$ is the MOND correction. This $\s\_M$ is then the dispersion {\it of the test-particle population whose scale height is being used.}
\par
The exact MOND correction depends on the MOND formulation assumed (e.g., Milgrom 2014b), and has to be calculated numerically for realistic discs. In the nonlinear Poisson theory of Bekenstein and Milgrom (1984), and in the QUMOND formulation (Milgrom 2010) used by Angus et al. (2015), the correction is given approximately (for thin discs) by
\beq  \z\_M(r)\approx1/\m[V^2(r)/r\az],  \eeqno{ii}
where $V$ is the local rotational speed, $\az$ is the MOND acceleration constant, and $\m(x)$ is the relevant MOND interpolating function (Milgrom 1989, 2001).
\par
Then, $\s\_M(r)$ are compared with the dispersion profile, $\s(r)$, measured and published as part of the DiskMass project.
\par
This is a test of MOND, different from, and independent of, rotation-curve analysis, although it is rather more dependent on the specific MOND theory assumed.
\par
The fault with MOND that Angus et al. claimed was as follows: If one uses in eq. (\ref{i}) $\Up\_K$ from the MOND rotation-curves analysis, and $h_z$ from $h_z-h\_R$ correlations for edge on galaxies, then the ratio
\beq \eta(r)\equiv\s\_M(r)/\s(r), \eeqno{iii}
which ideally should be 1 is, in fact, systematically larger than 1 (by about a factor of 1.4 on average).
\par
To me, the prime suspect cause for this discrepancy is the inadequacy of the DiskMass dispersions $\s$. I have suggested to Angus et al. that these are too low for the purpose they were used in the DiskMass applications and by Angus et al. (2015). This is because $\s$, which was derived from the line widths of integrated light, may be more strongly weighted by younger populations -- which have smaller dispersions -- while the K-band light distributed is more heavily weighted by old stars -- which have higher dispersions and larger scale heights. I also suggested to conduct a detailed analysis of the vertical column in the Milky Way near the sun, where the separation can be made, to assess the strength of this systematic error.
\par
Such an analysis has just been published by Aniyan et al. (2015), with results that support the above suspicion.
The correction factor they found for the solar column -- $\s$ being too small by a factor of $\sim0.7$ -- if it applies on average to the DiskMass galaxies, would bring the predicted MOND dispersions to remarkable agreement with the measured ones in both shape and magnitude.

In Sec. \ref{reev}, I discuss all these issues in more detail. Sec. \ref{disc} contains a brief general comment on possible lessons.

\section{\label{reev}Reevaluation}
Looking at the $\s\_M - \s$ comparisons of Angus et al. (2015), in their Figures 14-17, we note first that the predicted MOND profiles, match, rather well, the $\s(r)$ profiles {\it in shape}. In other words, $\eta(r)$ defined in eq. (\ref{iii}) is well consistent with being nearly $r$-independent, while $\s$ and $\s\_M$ separately are, generally, strongly $r$ dependent. For reasons discussed below, this, in itself, is {\it a highly nontrivial accomplishment of} MOND -- not discussed by Angus et al. (2015).
\par
\begin{figure}
\begin{center}
\includegraphics[width=0.8\columnwidth]{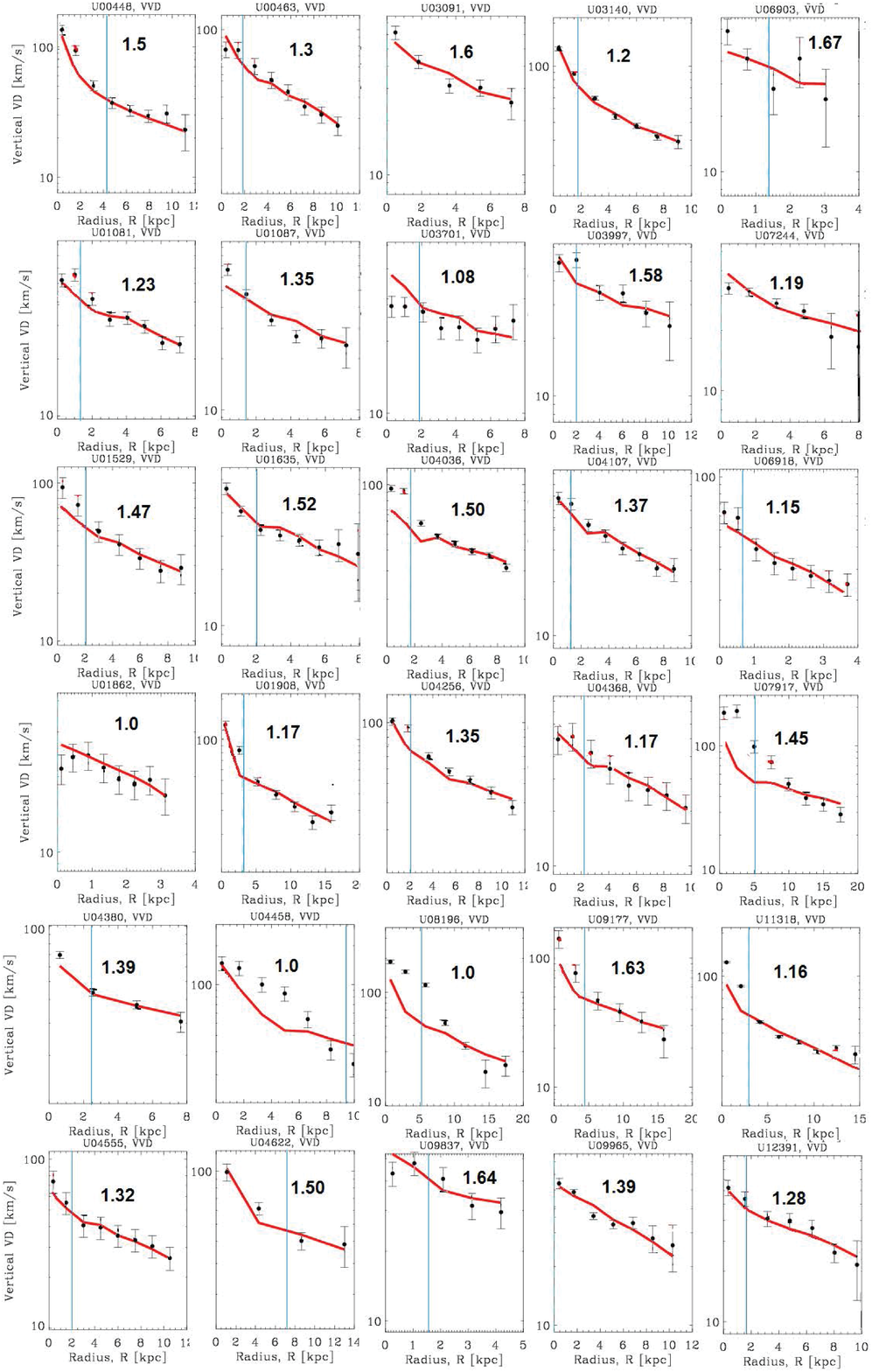}
\caption{MOND predictions $\s\_M(r)$, from Angus et al. (2015), (red line) and the DiskMass $\s(r)$ scaled up by a factor given in each frame for the 30 galaxies. The vertical turquoise line marks the bulge radius; data to its left are ``unreliable'', according to Angus et al.}\label{fig1}
\end{center}
\end{figure}

To bring this out more clearly, I plot in Figure \ref{fig1}, the MOND predictions, $\s\_M$, on top of the Diskmass data, corrected up by a factor, $\eta_c\s$, chosen for each galaxy to get a good match ``by eye''.
We see that, indeed, but for the overall scaling, MOND predicts the shapes of $\s(r)$ {\it very well}.
\par
The fault with MOND pointed by Angus et al. is then the fact that $\eta$, which supposedly should be 1, is systematically above 1. More quantitatively, I find from Figure \ref{fig1} that $\eta_c$ is between 1 and 1.7. It has a mean $\langle\eta\rangle\approx 1.34$, and a standard deviation $\s_\eta\approx 0.20$.
\par
Angus et al. also express this discrepancy [equivalently, as seen from eq. (\ref{i})] by stating that: ``... to match the DMS ($\s$) data, the vertical scaleheights would
have to be ... a factor of 2 lower than those derived from observations of edge-on galaxies with a similar scalelength.''
\par
The first culprit that came to my mind when first reading Angus et al. (2015) in preprint form, is well exposed in the following excerpts from my messages to them. For example (May 22 2015; a few misprints corrected):
``Regarding your recent paper: We know that stellar discs are made of various populations with very different z-dispersions and scale heights. Starting from the gas disc and young stars with low dispersion and small scale height to old populations with both rather much larger. When conducting z-analyses in the Milky Way one is always careful to consider a test-particle population for which  both the velocities and the z-distribution are used.

In your analysis (and that of the DiskMass people) all populations seem to be lumped together; i.e., the whole analysis is single component.

Is it not possible that the single velocity dispersion number you use is weighted differently by the different populations from the single number for the scale height. For example, if the dispersion is strongly weighted by young stars, but the light distribution from which the scale-height-scale-length relation is deduced (in the other, edge-on galaxies) is weighted by old stars you would get a mismatch of the type you find.
Shouldn't a proper z-analysis either be a multi-component one, or use the same population for both the distribution and the velocities of the test particles?
I looked only superficially, but you seem not to have discussed this issue.''
\par
And more (May 23): ``As I said, it only takes the sigmas you use to be 30 percent below the "correct" ones to produce a mismatch of a factor of 2 in the deduced scale height... Surely such small deviations could easily be produced by the procedure. The span in dispersions between different populations in the MW is much larger than this (I think about a factor 3)... One would have to take some model of a mixture of populations with their different dispersions, for example the MW populations, and look at what the line strengths are for the different types than integrate the line profiles and the photometry to see if we are comparing apples with oranges or not.
In any event your paper should clearly explain all this and in particular make the point that the factor of 2 mismatch in $h_z$ really is traced back to a small difference in sigma.''
\par
Angus et al. have not accepted this view, do not discuss these potential systematics, and do not question the adequacy of the DiskMass $\s$. Instead, they concluded: ``Thus, the most straight-forward way to reduce the model vertical
velocity dispersions in MOND is to decrease the stellar scaleheight.
If these were reduced from the values derived from observations
of edge-on galaxies by roughly a factor of 2 then the DMS vertical
velocity dispersions and rotation curves would be compatible
with MOND. Regardless of stellar scalelength, these disks would
have scaleheights between 200 and 400 pc. According to a two dimensional
KS tests, such thin disks are strongly at odds with
observations.''
\par
On the contrary, there is evidence to support the suspicion that the DiskMass $\s$ values are underestimates of what needs to be used in eq. (\ref{i}), and its Newtonian analogue, in conjunction with light distribution based on the $h_z-h\_R$ correlations and K-band photometry.
We see from Figure \ref{fig1} that we only need the DiskMass $\s$ to be 25 \% too small on average ($\eta=1.34$), to make the MOND predictions fully in agreement with the data.

If, on the other hand, we accept the DiskMass $\s$ as ``correct'' in the above sense, the following incongruities with other deductions appear:

(i) The resulting stellar $\Up\_K$ values that DiskMass collaboration deduced based on their $\s(r)$ (in Newtonian dynamics) are lower by a factor of $\sim 2$ than those deduced from population-synthesis. They are also substantially lower than values of $\Up\_K$ obtained by other means (see, e.g., McGaugh \& Schombert 2014 and Aniyan et al. 2015).

(ii) With these $\Up\_K$ values, galactic discs would be strongly submaximal. Namely, such discs would account for only a fraction of the observed rotational speeds at small radii -- even in high-surface-brightness, high-rotational-speed galaxies. This would mean, in Newtonian dynamics, that dark matter contributes substantially even near the centers of these galaxies. This is at odds with what is indicated by rotation-curve analysis (for example that features in the disc mass distributions are faithfully reproduced in the rotation curves, while they would have been washed out by a dark-matter halo).

(iii) Even more alarming, Angus, Gentile \& Famaey (2015) showed, more recently, that the procedure that has led to the published DiskMass $\Up\_K$ values was flawed, and, in fact, even overestimated $\Up\_K$. This is because even as they were, such low $\Up\_K$s imply the present of dark matter {\it in the disc itself}. Thus, some of the ``disc'' dynamical mass determined from $\s(r)$ has to be allotted to the dark matter in the disc. The deduced stellar masses, and $\Up\_K$, are thus even smaller. With this correction, the DiskMass $\s$s give $\Up\_K$ values a factor of $\sim 3$ lower than population-synthesis values. Recall that the MOND $\Up\_K$ values deduced from the rotation curves are consistent with population-synthesis values.
\par
Angus, Gentile \& Famaey (2015) list as one possible explanation of this disparity: ``Alternatively, the measured stellar velocity dispersions of the DiskMass survey might have been under-estimated by $>30\%$.''

(iv) With the low $\Up\_K$ values published by DiskMass, and even more so with the correction of Angus, Gentile \& Famaey (2015), the baryonic Tully-Fisher relation for predominantly-stellar disc galaxies would be substantially off that for gas-rich galaxies. All galaxies give a single, tight relation with $\Up\_K$ values consistent with population-synthesis results.

{\it All of these incongruities disappear if $\s$ given by DiskMass are corrected up by a factor $\sim 1.4$ on average (i.e., if they are underestimates by $\sim 30\%$ on average).}
\par
We can now understand why the fact that the MOND $\s\_M(r)$ match in shape the DiskMass $\s(r)$, in itself, very significant. Namely, why this is quite unexpected if the DikMass $\s$s, with their low $\Up\_K$s, are ``correct'', and dark matter were at play here, and not MOND.
\par
In the first place, the MOND correction, $\z\_M(r)$, is $r$- and galaxy-dependent. Why then, if MOND is not at action here, would $\z\_M(r)$ distort the shape of $\s\_M(r)$ in just the right way?
\par
Furthermore, in the dispersion plots of Angus et al. one does not reach regions where the rotational acceleration $V^2(r)/r\ll\az$, and so in these regions the MOND correction only reach modest values, perhaps, up to $\z\_M\sim 2-3$ in the outer regions studied for some of the galaxies.
In fact, the disagreement (in amplitude) between $\s\_M$ and $\s$ occurs also (and largely) in regions where $\z\_M\approx 1$. There, MOND predicts no discrepancy (i.e. a maximum disc situation), while the DiskMass $\s$s, with the small $\Up\_K$ values they imply, predict large discrepancies (highly sub-maximal discs) -- even larger with the caveat of Angus, Gentile \& Famaey (2015).
\par
If the latter were true, the dark-matter halo would have affected strongly both the rotation curve and the vertical disc dispersions; {\it and this in different and not directly related ways; in ways that are also $r$- and galaxy-dependent}: For example, the dark matter would enter rotation curves through its integrated mass up to the radius in question, but enter the dispersion essentially only through its local density within the disc. Also, the halo shape (e.g. ellipticity) enters the two in different ways (see, e.g., Angus, Gentile \& Famaey 2015), etc. Couple this with the fact, already mentioned, that MOND does introduce a galaxy- and $r$-dependent correction factor of up to $\sim 2-3$, it would be quite unexpected indeed if dark matter were responsible, and, with all the intricacies involved, MOND were to reproduce everything correctly, but for a constant correction factor for $\s$.
\par
{\it Indeed, correcting the DiskMass $\s$s in a way that eliminates the above-listed incongruities would also make the MOND predictions, $\s\_M(r)$, fully consistent with the observations, in (average) magnitude as well as in shape.}
\par
Now, Aniyan et al. (2015) have considered this issue of the inadequacy of velocity dispersions extracted from line profiles integrated over all populations. They suggest, indeed, that these may be underestimates relative to the representative ones, because they are weighted by young, cold populations that do not contribute much to the relevant light distribution.
\par
Aniyan et al. conducted a detailed analysis of data for the vertical column at the sun's position in the Milky Way, where it is possible to separate the different populations. They find that using the integrated line profile underestimates the ``correct'' dispersion -- that needed to conjoin dynamically with the light scale height -- by a factor of $\approx 0.7$. Similar results were gotten when the Besan\c{c}on model of the Galaxy was used.
\par
This is exactly the average correction needed to bring the predicted MOND $\s\_M(r)$ profiles of Angus et al.
to agreement with the DiskMass data, as it corresponds to $\eta\approx 1.4$.
This near coincidence may be somewhat fortuitous since in the sample studied by Angus et al. there is a range of $\eta$ values between 1 and 1.8.
And, the correction factor found by Aniyan et al. (2015) is probably not universal. It may depend or $r$ within a galaxy, and differ between galaxies, due to variation in population mix, details of the velocity distributions of different populations, etc. In addition, both $h_z$ and $\Up\_K$, which were assumed $r$-independent in the analysis, may, in fact, depend somewhat on $r$. On the other hand, the fact that $\eta(r)$ is found remarkably $r$-independent may be telling us that while the error correction factor one needs to apply to the DiskMass $\s$ may depend on the galaxy, the assumptions of $r$-independent $\Up\_K$ and $h\_z$ are approximately valid. These assumptions are also supported by direct observations of edge-on galaxies and studies of color variations across galaxies.
Aniyan et al. (2015) say they have embarked on a project to assess the correction factor at different radii, in external galaxies.
\par
Note also that the $h_z-h\_R$  correlation used by DiskMass and by Angus et al. for their adopted values of $h_z$ is said to have a standard deviation scatter of 25 \% in $h_z$. This in itself produces an error of $\sim 12\%$ in the predicted $\s\_M$, and can explain some of the scatter in $\eta$ that is found here.

\section{\label{disc}Discussion}
I conclude with another admonition I made to Angus et al. (May 23 2015) which carries a more general lesson:
``The general problem is, that after the more clear-cut tests of MOND, such as rotation curves, Tully-Fisher, etc., one is now starting to look at rather more subtle tests with only small, and more theory-dependent predicted discrepancies ($a\sim\az$), more roundabout astrophysical argumentations, with more iffy assumptions (as in your case). In these, one then has to be extra careful, as any small slip will lead to failure.''

\end{document}